\documentclass[manuscript]{acmart}
\renewcommand\footnotetextcopyrightpermission[1]{}
\usepackage{enumitem}

\copyrightyear{2026}
\acmYear{2026}
\setcopyright{cc}
\setcctype{by}
\acmConference[ICGJ 2026]{International Conference on Game Jams Hackathons and Game Creation Events}{August 21--22, 2026}{New York, NY, USA}
\acmBooktitle{International Conference on Game Jams Hackathons and Game Creation Events (ICGJ 2026), August 21--22, 2026, New York, NY, USA}
\acmDOI{10.1145/3833089.3833091}
\acmISBN{979-8-4007-2493-0/2026/08}

\begin{document}
\newcounter{RQCounter}
\newcommand{\RQ}[2]{%
  \refstepcounter{RQCounter}\label{#1}%
  \par\addvspace{4pt}%
  {\leftskip=4pt \rightskip=4pt \noindent
   \textbf{RQ}$_{\arabic{RQCounter}}$.~\emph{#2}\par}%
}
\newsavebox{\boxEcontent}
\newenvironment{boxE}
  {\par\addvspace{3pt}%
   \begin{lrbox}{\boxEcontent}%
   \begin{minipage}{\dimexpr\columnwidth-6pt\relax}}
  {\end{minipage}\end{lrbox}%
   \noindent
   {\setlength{\fboxrule}{0.6pt}\setlength{\fboxsep}{0.9pt}%
    \fbox{\textcolor{black!40}{\fbox{\textcolor{black}{\usebox{\boxEcontent}}}}}}%
   \par\addvspace{3pt}}

\newcommand{\hr}[1]{\textbf{RQ}$_{\ref{#1}}$}

\title{Quick Build, Careful Check? Generative AI Use in Hackathons}

\author{Wangyiyao Zhou}
\affiliation{
  \institution{Eindhoven University of Technology}
  \city{Eindhoven}
  \country{The Netherlands}
}
\email{w.zhou1@tue.nl}

\author{Alexander Serebrenik}
\affiliation{
  \institution{Eindhoven University of Technology}
  \city{Eindhoven}
  \country{The Netherlands}
}
\email{a.serebrenik@tue.nl}

\author{Alexander Nolte}
\affiliation{
  \institution{Eindhoven University of Technology}
  \city{Eindhoven}
  \country{The Netherlands}
}
\affiliation{
  \institution{Carnegie Mellon University}
  \city{Pittsburgh}
  \state{PA}
  \country{USA}
}
\email{a.u.nolte@tue.nl}

\begin{abstract}
Hackathons are time-bounded events where participants form teams to rapidly build software projects. Their short-term nature makes them a natural setting for Generative AI (GenAI) use, given its promise of speed and efficiency. Yet despite GenAI being increasingly adopted in these events, we still know little about what participants use GenAI for and, crucially, what they do not use it for and why. We report an exploratory interview study with participants from different teams at a two-day AI-themed hackathon in central Europe. We found that participants used GenAI for purposes beyond coding, including learning unfamiliar topics, brainstorming, and preparing documentation. At the same time, they combined multiple GenAI and non-GenAI tools depending on task fit. We also found that, despite the absence of team rules, all participants we studied converged on an unwritten practice of checking GenAI output before using it, yet their ability to actually verify was constrained by time pressure and limited domain knowledge. This work aims to identify avenues for further investigation by outlining a follow-up study combining observation, surveys, and prompt-and-response logs.

\end{abstract}

\keywords{Hackathons; Generative AI; Verification; Software Engineering}
\begin{CCSXML}
<ccs2012>
   <concept>
       <concept_id>10003120.10003130.10011762</concept_id>
       <concept_desc>Human-centered computing~Empirical studies in collaborative and social computing</concept_desc>
       <concept_significance>500</concept_significance>
       </concept>
 </ccs2012>
\end{CCSXML}

\ccsdesc[500]{Human-centered computing~Empirical studies in collaborative and social computing}
\maketitle

\section{Introduction}
GenAI tools such as ChatGPT and GitHub Copilot have become increasingly integrated into software engineering work~\cite{nguyenDuc2025genaiAgenda}. Studies report measurable productivity gains from coding assistants~\cite{cui2025effects, mohamed2025impact}, and also show that developers use GenAI not only to generate code but also to learn unfamiliar topics and get guidance on how to approach tasks~\cite{khojah2024beyond}. 
Hackathons are time-bounded collaborative events~\cite{falk2024future} in which participants rapidly design or build an artifact~\cite{OnHackathonsSLR}, often with software as a central outcome~\cite{Pe-Than2019Corporate}. In this setting we expect GenAI to be used extensively, since the efficiency gains reported in everyday software work~\cite{cui2025effects} matter most when time is heavily constrained. Participants are likely to use a variety of tools, and to use them in different ways depending on their tasks and prior experience. Existing work has begun to describe GenAI use in hackathons through event-level surveys~\cite{sajja2024integrating} and observations of specific groups such as novice programmers~\cite{gama2025vibes}. Each approach has complementary limits: surveys offer breadth but limited depth, while observation captures behavior but not the perceptions and decision making behind it. Interviews are better suited to participant perspectives, allowing us to understand how participants reason about their choices, yet this angle has not been applied to how teams in hackathons choose tools and divide tasks with GenAI. To address this gap we ask:

\RQ{RQ1}{How do participants in hackathon teams use GenAI across different tasks and tools during the event?}

\noindent Beyond what GenAI is used for, how its output is checked is also an open question. Prior work shows that time pressure leads developers to deprioritize testing and quality assurance in everyday software work~\cite{kuutila2020timepressure}. It is therefore plausible that hackathon participants readily accept GenAI output, deprioritizing verification compared to their everyday use. There is existing work on how software communities~\cite{yang2026beyond} and knowledge workers~\cite{wagman2025genai} govern GenAI use, but how participants in hackathon teams handle this under time pressure is not yet understood. We therefore ask:

\RQ{RQ2}{How do participants verify GenAI outputs under hackathon time pressure?}

\noindent To address these questions, we conducted an exploratory qualitative interview study with participants from different teams at a two-day AI-themed hackathon. We found that participants combine multiple GenAI and non-GenAI tools by task fit rather than by team policy, converge on an unwritten practice of checking GenAI output even without explicit team rules, and face verification constraints from time pressure and limited domain knowledge.

\section{Related Work}
Hackathons have been studied across a wide range of contexts, including industry product innovation~\cite{Komssi2015WhatAreHackathonFor}, corporate engineering~\cite{Pe-Than2019Corporate}, and software engineering education~\cite{Porras2018HackathonsEdu}. Before GenAI, hackathon teams already had to coordinate ad hoc design and decision making under acute time pressure, often relying on informal design thinking practices~\cite{gama2023developers}. Two features make hackathons a distinctive site for studying how participants work with new tools: the compressed schedule forces rapid tool and workflow decisions, and teams often consist of individuals who have not collaborated before, so practices around tool use and verification are negotiated on the spot rather than inherited from a stable team culture~\cite{peThan2022corporate}. Our study extends this work by asking how teams choose tools, divide tasks with GenAI, and check its outputs under time pressure.

\looseness=-1 Outside hackathons, an emerging body of empirical research discusses how developers integrate GenAI into everyday work. Existing studies found that practitioners use ChatGPT for guidance and to learn about topics, rather than to receive ready-to-use code~\cite{khojah2024beyond}. Studies also report that software professionals treat GenAI as an assistive tool rather than a standalone solution, with persistent concerns about output accuracy~\cite{santos2025modelassisted}. A systematic review of 39 studies finds that GenAI assistance accelerates development but raises concerns about cognitive offloading and code quality~\cite{mohamed2025impact}. These studies establish that verifying GenAI output is a routine concern in everyday development. Beyond individual practice, parallel work has begun to examine how GenAI use is governed at the level of software communities~\cite{yang2026beyond} and knowledge work organizations~\cite{wagman2025genai}. 

Research at this intersection is still emerging and touches on three dimensions: tool combinations, task allocation, and verification. Sajja et al.~\cite{sajja2024integrating} surveyed 151 participants of an educational hackathon and reported that GenAI accelerated coding and brainstorming but raised concerns about over-reliance and ethical implications. Gama et al.~\cite{gama2025vibes} observed novice programmers engaged in vibe coding during a one-day hackathon and found that teams combined multiple AI tools in pipeline configurations while human judgment remained essential, especially given uneven code quality that required rework. Li et al.~\cite{li2026hackagents} took a more interventionist approach, designing a phase-specific multi-agent system that differentiates AI assistance across ideation, implementation, and evaluation, deployed in two innovation hackathons to mitigate over-reliance and preserve idea diversity.

Taken together, these studies indicate the proliferation of GenAI in hackathons and point to both benefits and tensions. Existing studies, however, either aggregate across many participants, which allows broad overviews but limited insight into how specific teams choose tools and coordinate work, or focus on different aspects such as novice engagement or tool design. What remains absent is an in-depth account of how participants in different teams combine tasks, tools, and verification practices in a single hackathon, and where the limits of these practices lie.

\section{Method}
To answer our research questions, we conducted an exploratory qualitative study using semi-structured interviews~\cite{hove2005experiences}, a method suited to capturing participants' reasoning and perspectives on their GenAI use, which fixed-response data would miss.

\noindent\textbf{Event setting.} 
We studied a two-day AI-themed hackathon held in central Europe in September 2025, with 90 participants across 10 teams. Projects were judged on several criteria, one of which was the skillful use of AI. The setting suited our questions: the 28-hour format produced the time pressure relevant to \hr{RQ2}; the AI evaluation criterion primed participants to articulate their GenAI choices; and participants could choose freely among the available tools, including a featured open-source LLM and commercial assistants, relevant to \hr{RQ1}.

\noindent\textbf{Participants.} We sampled along two dimensions: (i) one participant per team, to capture cross-team variation in GenAI use; and (ii) a mix of professional backgrounds, including students and software engineers, since background plausibly shapes how participants approach tool choice and verification. Table~\ref{tab:participants} summarizes the four participants and their team projects. All four reported prior GenAI use. Interviews were conducted by the first author, who had no prior relationship with the participants. Participation was voluntary, and the study was approved by our institutional Ethical Review Board (ERB-338).

\begingroup
\setlength{\tabcolsep}{1.5pt}
\setlength{\textfloatsep}{4pt plus 1pt minus 1pt}
\setlength{\intextsep}{4pt plus 1pt minus 1pt}

\begin{table}[!htbp]
\vspace{-2pt}
\captionsetup{skip=2pt}
\caption{Participant overview}
\label{tab:participants}
\centering
\footnotesize
\renewcommand{\arraystretch}{1.05}

\begin{tabular}{@{}p{0.06\columnwidth} p{0.36\columnwidth} p{0.52\columnwidth}@{}}
\hline
\textbf{ID} & \textbf{Background} & \textbf{Project} \\
\hline
P1 & Student in CS & Book image matching \\
P2 & Software engineer & Synthetic solar systems \\
P3 & Student in robotics & Farm logistics app \\
P4 & Software engineer & Low resource language chatbot \\
\hline
\end{tabular}
\vspace{-2pt}
\end{table}
\endgroup

\noindent\textbf{Interviews.} Each interview lasted about 30 minutes and followed a semi-structured guide with three blocks. The first covered prior experience with hackathons and GenAI, providing a baseline for distinguishing hackathon-specific choices from prior habits. The second opened with the team's project goals, then asked where GenAI was helpful or unhelpful and whether teams had rules for its use. The third invited reflection on downsides and willingness to use GenAI again, surfacing participants' retrospective views on verification. The full guide is in the Appendix~\ref{app:guide}.

\noindent\textbf{Analysis.} The first author analyzed the interview transcripts using thematic analysis~\cite{braun2006using}, with transcripts imported into ATLAS.ti and coded iteratively. Initial codes were generated close to the data, capturing concrete actions, tools, and reflections; examples include \textit{generate code} and \textit{tool: Claude} relevant to \hr{RQ1}, and \textit{line-by-line review} and \textit{limited ability to verify} relevant to \hr{RQ2}. Through repeated reading and discussion with the coauthors, related codes were grouped into three themes: task allocation, combining tools by task fit, and verification practice.

\section{Findings}
Sections~\ref{sec:org} and \ref{sec:tool} address \hr{RQ1} by describing what participants used GenAI for and how they combined it with other tools. Section~\ref{sec:verifying} addresses \hr{RQ2} by describing how participants checked GenAI output under time pressure.

\subsection{Where GenAI was used and where not}
\label{sec:org}
Across the four cases, participants used GenAI for four kinds of work, discussed below: entering new topics, ideation and planning, code and implementation, and documentation and presentations.

\noindent\textbf{Entering new topics. }All four participants used GenAI to engage with topics or techniques outside their prior expertise. P1 started two days before the event, working through an unfamiliar paper with ``so many terminologies that I did not understand'' (P1). P1 also saw GenAI as a judgment-free conversational partner: ``I ask simple or ‘stupid’ questions, and through that, I deepen my understanding before I actually start working on the topic'' (P1). P3 used GenAI only after reading other sources: ``For me it was more efficient to read a book or something and then ask the AI specific questions to get a specific answer'' (P3). P4 learned NLP while building a low-resource language chatbot, describing the hackathon as the ``first time I gained so much experience'' (P4) in NLP, partly through AI and partly through a teammate with a stronger technical background. P2, a software engineer, worked in an unfamiliar data science domain throughout the event, relying on a GenAI-integrated Jupyter environment to navigate it: ``I don't have a strong data science or mathematical background'' (P2).

\noindent\textbf{Ideation and planning.} The two students in our sample (P1, P3) used GenAI to generate ideas and plan teamwork. Before the event, P1 used GenAI to find research directions: ``I asked ChatGPT to give me some ideas of what people use'' (P1). During the event, P3's team used GenAI to help split tasks and, as P3 put it, ``for ideas and brainstorming [...] turned out quite good'' (P3). In both cases, GenAI was a starting point rather than a decision maker.

\noindent\textbf{Code and implementation. }All four participants used GenAI to speed up coding under time pressure. P1 used GenAI ``a lot to help generate the basic code'' (P1). P3 found GenAI ``quite helpful'' (P3) for small, self-contained code: ``if you have a function, if you have a loop or something, there it was quite helpful'' (P3). P4 said plainly: ``without AI, we could not have finished this project in two days'' (P4). P2's coding was mediated by an AI-integrated notebook environment that could ``write code and execute it'' (P2). P1 also drew a boundary by data type: ``If the data was potentially sensitive, then of course we wouldn't use AI for that'' (P1). P3 drew another boundary, by who would read the code afterwards: ``if you give AI a code snippet with comments documented in a style your team understands, it might change it. Maybe the comments are gone, or they are completely changed, and then you do not understand anything and the knowledge is basically lost'' (P3). Across participants, GenAI was perceived as useful for self-contained snippets but risky for code others would later read or modify.

\noindent\textbf{Documentation and presentations. }Three participants (P1, P3, P4) used GenAI for text deliverables such as documentation, academic writing, and presentation slides. P1 used GenAI to refine academic writing: ``I write something in my own words, and then I ask it to rephrase or improve it'' (P1). P3 said GenAI was useful for correcting documentation quickly when time was short (P3). A teammate of P4 generated a full slide deck from a short outline: ``you just write the outline, the idea, what the project is [...] and the AI can generate the whole presentation for you'' (P4). Yet participants also intentionally set boundaries. P2's team kept the presentation manual: ``The presentation was mostly done manually'' (P2). P4 deliberately reserved design work for humans: ``we still need human creativity [...] AI can also design, but humans have a different artistic sense'' (P4).

\subsection{Combining tools by task fit}
\label{sec:tool}
Participants did not treat GenAI as a single, general-purpose tool. They described different AI systems as having different strengths and chose between them accordingly. P4 made this explicit: ``Claude is mostly for engineering, but OpenAI is very useful for the business model'' (P4). P3's team started the event by trying to rely only on the event's featured LLM, but moved to other tools after finding that ``it had this hallucination issue, and sometimes it actually produced incorrect code'' (P3). P2 had no default tool before the event: ``I did not know which challenge I would be part of'' (P2), and ``spontaneously searched for AI tools that work well with data science tasks'' (P2) once the challenge was clear, ending up with Deepnote (a Jupyter notebook with built-in AI assistant).

The participants mentioned tools that can roughly be structured into three categories. The most visible was chat-based GenAI assistants, used by P1, P3, and P4 (ChatGPT, Claude, Copilot, Perplexity, DeepSeek, and the event's featured LLM). A second layer was tools with GenAI built in. P2 used Deepnote (above); P4 used Cursor, noting that ``the AI agent helps me with development. It can also check and test everything'' (P4). A third layer was non-GenAI tools, which GenAI did not replace. P1 first ``saw something on Stack Overflow'' (P1), then asked ChatGPT for directions about the field, then searched Google Scholar: ``I used those suggestions to search on Google Scholar for other researchers'' (P1).

Tool choice mixed prior habit with on-the-fly adaptation: P1 and P3 leaned on their everyday assistants, while P2 and P4 picked up new platforms (Deepnote, Cursor) when tasks demanded them. Tool fit was also shaped by how code would travel within a team.

\subsection{Verifying AI output under time pressure}
\label{sec:verifying}
No team set explicit rules for GenAI use, and every participant used GenAI individually. P1: ``everyone used AI if they felt like it. There were no limitations'' (P1). P3 elaborated: ``Everyone has their own GitHub account and uses their own AI [...] we do not share accounts, and we do not really have rules for AI usage'' (P3).

Yet despite this absence of shared rules, every participant described wanting to check or adjust GenAI output before using it. P1: ``I used AI a lot to help generate the basic code, which I then refactored'' (P1).  P3 described his team's practice as a collective response to repeated hallucinations: ``we never really took an answer from the AI and used it directly. We always edited it and changed it, especially for code'' (P3). P4: ``we still have to check it, line by line, because sometimes the generated code is not correct'' (P4). For P4's team, verification took the form of model cross-checking: ``we relied a bit on [the event's featured LLM], but at the same time we also checked with another model like DeepSeek'' (P4). Two participants explained why they checked. P1 framed it as individual responsibility: ``the person has to be able to explain and take responsibility for what they're suggesting, even if it came from an AI'' (P1). P3 framed it at team level: ``it is more like group pressure to not let all of your code be written by AI [...] you still have to understand the code'' (P3). Both students shared the same expectation: being able to understand and account for AI-generated code.

\looseness=-1 But wanting to check did not always mean being able to. Two constraints recurred across the cases. The first was domain knowledge. P2, a software engineer, working in unfamiliar data science: ``I often didn't really understand whether it was doing the right thing or not, because I'm not that strong mathematically'' (P2). P2 contrasted this with familiar-domain projects: ``I can see when it's hallucinating'' (P2). In data science, P2 ``couldn't really judge a lot of the time'' (P2). The second constraint was time. Outside hackathons, P1 described a reviewable workflow: ``I try to give smaller chunks of tasks to the AI. This way I can reinterpret its output and incorporate it into my code'' (P1). The hackathon changed this: ``during the hackathon, I had to let it write a bigger chunk of code'' (P1). Afterwards, P1 reflected: ``I would try not to use it as extensively as I did during the last hackathon. There is something in me that is afraid I will not be able to explain every little thing'' (P1). 

Beyond the immediate constraints, P1 reflected on a longer-term cost of relying on AI under time pressure. P1 described how using AI ``contributes to impostor syndrome. Even though I understand what I am doing, I cannot replicate it myself so fast, and it makes me doubt myself when I apply for positions and jobs'' (P1). For P1, the gap between using AI-generated code and being able to reproduce it independently thus extended past the event into doubts about professional competence. 

\begin{boxE}
\textbf{Key findings.}
\begin{itemize}[nosep,leftmargin=*]
  \item Participants used GenAI for learning, ideation, coding, and documentation, and set task-specific boundaries.
  \item Participants combined chatbots, AI-embedded tools, and traditional sources by task fit.
  \item Despite no explicit team rules, participants checked GenAI output, constrained by time pressure and domain knowledge.
\end{itemize}
\end{boxE}

\section{Discussion}
We identified similarities to prior work in relation to how participants used GenAI across tasks~\cite{khojah2024beyond} and combined it with other tools~\cite{gama2025vibes} (\hr{RQ1}). We also found different strategies for verification (\hr{RQ2}), which we discuss in more detail in the following.

\subsection{Checking emerges without rules}
None of the teams had explicit rules for GenAI use, and as far as participants reported, the expectation to check GenAI output was never taught, discussed by organizers or mentors, or written down. Yet all four participants described this expectation as something they held, framing it either as individual responsibility or as team-level pressure to still understand one's own code. Prior hackathon studies have framed over-reliance as something the absence of rules invites~\cite{sajja2024integrating} or that needs structured interventions~\cite{li2026hackagents}. Our cases suggest a different starting point: participants brought a verification habit from their everyday work into the hackathon. This contrasts with software communities and larger organizations, where similar concerns are increasingly addressed through explicit governance~\cite{yang2026beyond, wagman2025genai}, and complements calls for explicit scaffolds for critical evaluation of AI outputs~\cite{gama2025vibes} by showing that participants already develop informal versions of such evaluation.

The presence of an informal checking practice shifts the question for organizers from whether to permit GenAI use to how to support the checking practices participants were already trying to sustain in the cases we studied. One concrete direction could be to treat explainability of contribution, that is, being able to account for what one submitted, as an evaluation dimension alongside skillful use of GenAI. This evaluation criterion parallels proposals in programming education, where Ye et al.~\cite{ye2026attribution} argue for process-oriented attribution that requires students to demonstrate engagement with AI rather than only disclose its use.

\subsection{Time pressure constrains verification}
Participants who held themselves to a checking practice also described not always being able to meet it. Two constraints appeared. The first was capability: in domains they did not fully understand, they could not reliably judge whether AI output was correct. This mirrors how a lack of domain knowledge has long been recognized as a barrier to contribution in software projects~\cite{steinmacher2014barriers}, now resurfacing as a barrier to verifying GenAI output. The second was time: participants who preferred small reviewable chunks abandoned that habit under pressure and accepted larger, less scrutinized contributions. This echoes Afroz et al.~\cite{afroz2026fastspurious}, who report that speed gains from GenAI are offset by increased review burden and verification load in everyday software work. In hackathons the trade-off is sharpened: verification competes for the same limited time as building, debugging, and presenting, and appears to be among the steps most likely to be cut. This is not an argument against using GenAI in hackathons, but a reason to plan for verification as part of the work. Constrained verification may affect not only output correctness but also participants' confidence in their own competence, connecting P1's account to impostor feelings documented among software engineers~\cite{guenes2024impostor}. The tension is likely to sharpen as participants adopt more agentic AI tools, which take on larger and more autonomous units of work and thus leave less for the human to review, as P4 was already doing with Cursor.

\section{Limitations}
Our study has four main limitations. First, the sample is small (four participants from one event), which is appropriate for an exploratory study aimed at surfacing aspects worth further inquiry rather than generalizable claims. Second, recruitment introduced selection bias: participants self-selected through the organizers, and studying different participants might have yielded different observations. We also interviewed only one member per team, which means our findings rely on a single perspective per team and other team members might have perceived things differently. We treat our cases as accounts from engaged users rather than from a representative cross-section, and read team-level claims (such as the absence of GenAI rules) as one member's perception rather than a shared perception across the team. Third, our data are perception-based; participants discussed how they perceived using GenAI, which might be different from their actual behavior. We accept this limitation because our questions concern participants' reasoning about GenAI use, which interviews are well suited to capture. Fourth, the transcripts were coded by a single researcher; we mitigated this through repeated discussion with coauthors~\cite{lincoln1985naturalistic}.

\section{Conclusion and Future Work}
We studied how participants across four hackathon teams used and verified GenAI outputs. GenAI aided learning, ideation, coding, and documentation; verification emerged informally even without explicit rules, while time pressure and limited domain knowledge constrained how well participants could check outputs. Future work should broaden the sample across multiple hackathons and move beyond self-report through observation, surveys, and prompt-and-response logs, with particular attention to agentic tools that delegate larger units of work to AI, to participants' motivations, and to whether GenAI merely seeds ideas or shapes final decisions.

\bibliographystyle{ACM-Reference-Format}
\bibliography{references}

\appendix
\section{Interview Guide}
\label{app:guide}

\small

\noindent\textbf{Introduction} This interview is part of a research project about how people use Generative AI tools like ChatGPT, Gemini, or image generators like Midjourney during hackathons. We are interested in what you expected before the event, how you and your team used these tools during the hackathon, and what you thought about the experience afterwards. The interview will be audio recorded, but your name will not be used and everything will be pseudonymized. You can skip any question or stop the interview at any time.

\medskip
\noindent\textbf{Section 1: Expectations before using GenAI}
\begin{enumerate}[nosep,leftmargin=*,topsep=2pt,itemsep=1pt,parsep=0pt]
  \item Can you tell me a little about your background?
  \item Can you think of a situation where you used AI before? Which AI tools have you used, and for what? \textit{Probe:} Have you used AI in a collaborative setting before?
  \item Was this your first hackathon? How many have you participated in?
\end{enumerate}

\medskip
\noindent\textbf{Section 2: Actual use during the hackathon}
\begin{enumerate}[nosep,leftmargin=*,topsep=2pt,itemsep=1pt,parsep=0pt]\setcounter{enumi}{3}
  \item What made you join this hackathon? Which team were you part of?
  \item What did your team plan to do during the hackathon?
  \item Was there a situation where AI was especially helpful? \textit{Probe:} Which tool? What did it help with? Who in the team used it?
  \item Was there a situation where AI was not helpful, or even caused problems? How did you or your team deal with it?
  \item Did your team discuss when or how to use AI? Were there any shared rules?
  \item What did your team decide \textit{not} to use AI for?
\end{enumerate}

\medskip
\noindent\textbf{Section 3: Perceptions and reflections}
\begin{enumerate}[nosep,leftmargin=*,topsep=2pt,itemsep=1pt,parsep=0pt]\setcounter{enumi}{9}
  \item How would you describe your overall experience using AI during the hackathon?
  \item Compared to your expectations before the event, how was the actual experience?
  \item Was your experience using AI in the hackathon different from your prior use in school, work, or personal projects?
  \item Were there any downsides or challenges, such as misunderstandings, incorrect outputs, or disagreement?
  \item Did you try any AI tools you had not used before?
  \item Did you give credit to AI in your final submission or presentation?
  \item Would you use AI again in future projects?
\end{enumerate}
\normalsize

\end{document}